\documentclass[12pt]{article}

\usepackage{amsmath, amssymb, amsfonts}
\usepackage{graphicx}
\usepackage[numbers,super]{natbib}
\usepackage[colorlinks=true, linkcolor=blue, citecolor=blue, urlcolor=blue]{hyperref}
\usepackage{caption}
\usepackage{float} 
\usepackage{geometry}
\usepackage{physics} 
\usepackage{doi} 

\makeatletter
\renewcommand\@biblabel[1]{#1.}
\makeatother

\title{\centering Algorithmic Advantage on a Gate-Based Photonic Quantum Neural Network}

\author{
  Solomon McKiernan$^{1,2}$\thanks{\texttt{srm201@cam.ac.uk}},
  Luca Sapienza$^{1}$\thanks{\texttt{ls2052@cam.ac.uk}} \\
  \small $^{1}$Department of Engineering, University of Cambridge, Cambridge CB3 0FA, United Kingdom \\
  \small $^{2}$Leonardo UK Ltd, Luton LU1 3PG, United Kingdom
}

\date{} 

\begin{document}
\maketitle


\begin{abstract}
We report on a gate-based variational quantum classifier implemented with single photons and probabilistic gates, to emulate the standard quantum circuit model framework. We evaluate the expressive power of two deployable quantum neural networks (QNNs) by computing their effective dimension, a capacity measure grounded in a proven generalization-error bound, and compare them with classical artificial neural networks (ANNs) of equivalent trainable-parameter count. Supervised binary classification tasks are used to benchmark performance across photonic and superconducting QNNs, both of which exhibit superior converged (lower) cross-entropy loss and (higher) prediction accuracy relative to matched-parameter ANNs. For a nonlinearly separable task, our photonic QNN with a single pair of trainable parameters successfully converged (loss 0.04 and accuracy 100\%), whereas the equivalent ANN failed to learn the decision boundary, saturating at random-guessing performance. We simulate photonic quantum circuits, training them on the XOR problem and a two-class Iris subset using gradient-free optimization, and assess their robustness to sampling errors under realistic noise processes including photon loss and phase-shifter imperfections. Circuits with comparatively high effective dimension were deployed remotely on a six-qubit photonic quantum processor, achieving classification accuracies of up to 100\% in both online and offline learning settings. Notably, even the simplest QNN deployed, with just two trainable parameters, successfully solved tasks that classically require ANNs with at least quadruple the number of parameters, suggesting an emergent algorithmic advantage. Overall, these results demonstrate a clear proof-of-principle that gate-based QNNs can be realized and trained effectively on current photonic hardware, providing proof of algorithmic advantage on a gate-based photonic QNN.
\end{abstract}

Traditional machine learning models, such as feedforward artificial neural networks (ANNs), have enabled significant advances in tasks including image and signal classification, pattern recognition, and anomaly detection \cite{annPatRecAbiodun2019}. However, their training cost increases rapidly with the size and complexity of the data set, especially as data-intensive applications, such as large-scale internet traffic analysis, continue to expand \cite{deepLearnNajafabadi2015}. Quantum machine learning (QML) seeks to address these challenges by combining quantum computing with classical learning techniques. QML algorithms can learn directly from quantum data and have been experimentally shown to be advantageous over classical approaches \cite{quantumDataHuang2022}. However, more common applications involve classical datasets, and although QML has been analytically shown to offer exponential speedup for classification tasks in specific cases \cite{robustQmlSpeedupLiu2021}, achieving such advantages generally requires large fault-tolerant quantum computers, which might remain decades away \cite{qTimelineMosca2024}.

Photonic quantum processors are a promising platform for practical near-term QML \cite{ascellaMaring2024}. Single-photon architectures enable native implementations of quantum neural networks (QNNs) while benefiting from low decoherence, modest cooling requirements, and natural compatibility with both classical and quantum communication networks, with recent work experimentally demonstrating a variational classifier using photon-native operations \cite{ascellaMaring2024}. However, these demonstrations do not directly correspond to gate-based QNN models, such as those analyzed in numerical studies by Abbas et al., who reported enhanced effective dimension (ED) and faster training relative to comparable ANNs \cite{qnnAbbas2021}. Bridging this gap between numerical gate-based results and experimental single-photon hardware remains an outstanding challenge.

\begin{figure}[H]
    \centering
    \includegraphics[width=\textwidth]{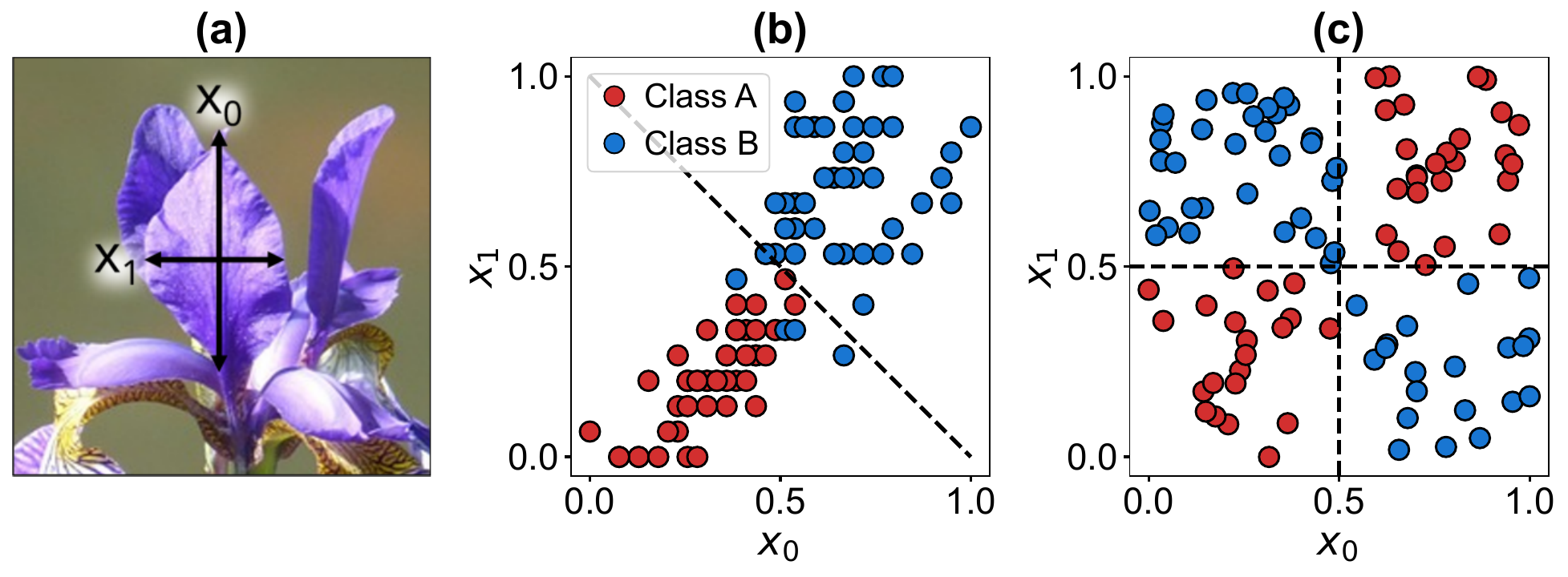}
    \caption{(a) Example labeled image of an Iris flower, showing features $x_0$ (petal length) and $x_1$ (petal width). (b) Nearly linearly separable Iris subset (Versicolor = Class A, Virginica = Class B), with feature lengths normalized to 1. (c) Nonlinearly separable XOR dataset, with Class A corresponding to the bottom-left and top-right clusters and Class B to the top-left and bottom-right clusters. Dashed lines indicate illustrative approximate decision boundaries highlighting the separability structure of each dataset.}
    \label{fig:inputData}
\end{figure}

Here, we address this gap by experimentally implementing a gate-based variational quantum classifier using single photons and probabilistic linear-optical gates. We evaluate the expressive power of deployable photonic QNNs through their effective dimension and benchmark them against classical ANNs with matched parameter counts (i.e., the same number of classically optimized trainable weights), as well as against a superconducting-qubit QNN. To test learning performance, we consider two binary-classification datasets: the first is a pseudo-randomly generated XOR dataset, a canonical nonlinearly separable problem that requires a multilayer classical ANN to solve \cite{minXorSingh2016}; the second is a nearly linearly separable two-class subset of the Iris dataset
as provided by scikit-learn \cite{sklearnPedregosa2011}, chosen to contrast performance on a simpler decision boundary and to enable comparison with a single-layer classical ANN (i.e. a perceptron). These datasets, shown in Fig.~\ref{fig:inputData}, provide complementary benchmarks for assessing the expressive capacity and trainability of photonic gate-based QNNs under realistic hardware constraints.

Following the Knill–Laflamme–Milburn (KLM) scheme for linear-optical quantum computation \cite{KLM2001}, the Ascella photonic processor used in this work implements probabilistic gate-based photonic circuits using dual-rail qubits and reconfigurable interferometers \cite{ascellaMaring2024}. Ascella allows for single-qubit rotations with at least one CNOT gate (two-qubit entangling operation) within the chip’s 12-mode architecture \cite{ascellaMaring2024}. This hardware constraint limits the depth of implementable variational circuits but still allows the construction of small ($\leq 6$ qubits), expressive ($>0.9$ effective dimension) QNNs. These QNNs adhere to the standard definition of variational quantum classifiers: trainable quantum circuits built from parameterized qubit rotations, whose angles are classically optimized to predict class labels from encoded data \cite{qnnAbbas2021, circuitQnnSchuld2020}.

\begin{figure}[H]
    \centering
    \includegraphics[width=\textwidth]{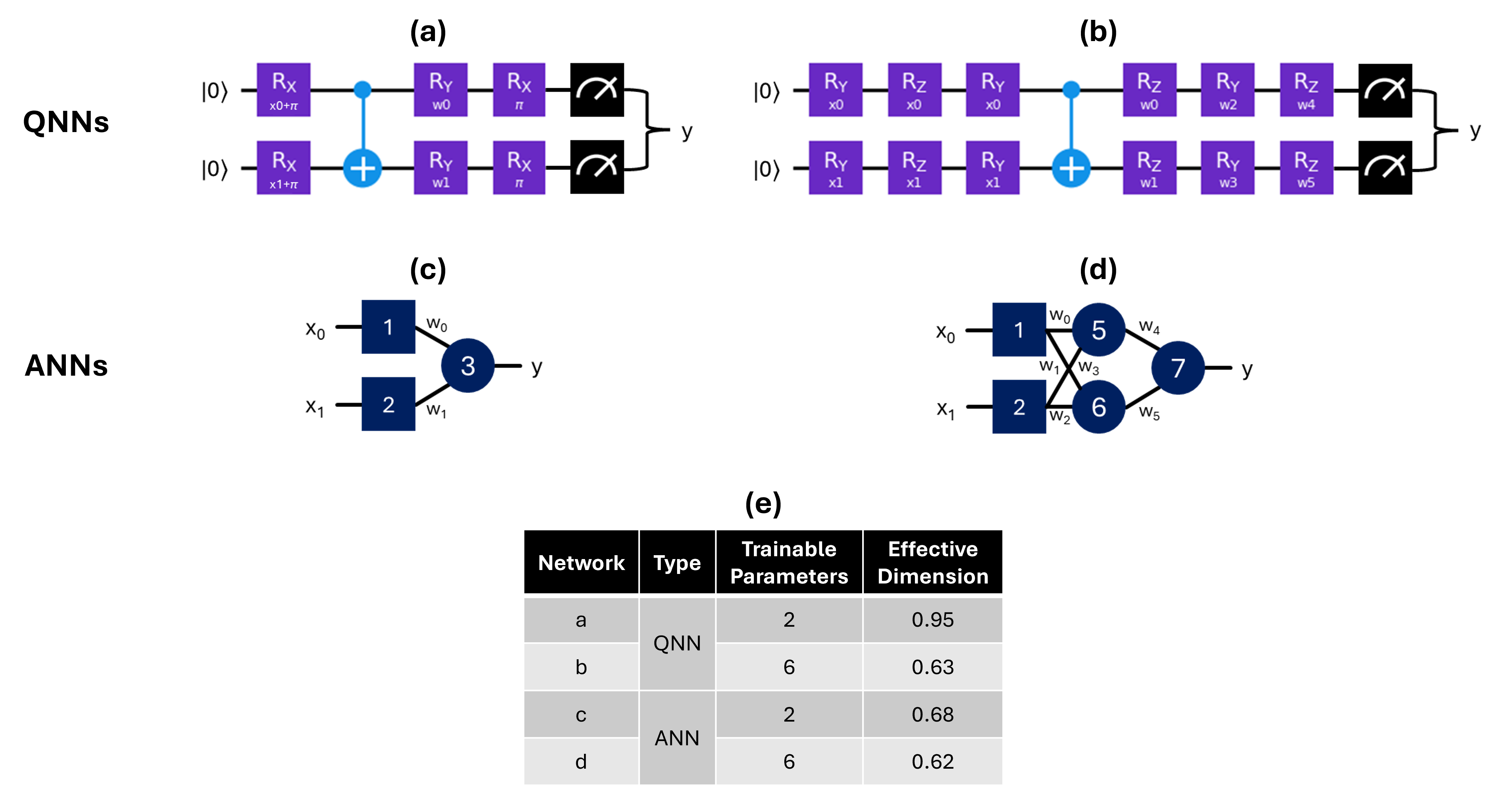}
    \caption{Classical and quantum neural network architectures with $n+1$ trainable parameters ${w_n}$ and corresponding effective dimensions. (a, b) Two-qubit variational quantum circuits using angle encoding, parameterized single-qubit rotations ($R_x$, $R_y$, $R_z$), a single CNOT entangling gate, and measurement-based estimation of the output $y$. (a) Two-parameter QNN. (b) Six-parameter QNN. (c) Two-parameter ANN: a single-layer perceptron with two trainable weights. (d) Six-parameter ANN: a multilayer perceptron with one hidden layer and six trainable weights, enabling nonlinear decision boundaries. (e) Converged normalized effective dimensions.}
    \label{fig:qnnCircuits}
\end{figure}

We compare two classical ANNs with two QNNs, with all four models shown in Fig.~\ref{fig:qnnCircuits}. Each pair consists of a minimal model with two trainable parameters and a deeper model with six trainable parameters, allowing a fair comparison between the ANNs and QNNs since both use the same optimizer, loss function, and parameter count. The ANN architectures are fully connected feedforward networks with two input neurons matching the two input data features. Similarly, the QNNs have two qubits to encode these two classical inputs. All four models are trained on identical datasets (Fig.~\ref{fig:inputData}) for direct comparison of representational capacity.

Quantum circuits are initialized with each qubit in the computational basis state $\ket{0}$. In the dual-rail photonic encoding used here, $\ket{0}$ corresponds to a single photon in the upper rail (waveguide), while $\ket{1}$ corresponds to a photon in the lower rail. Parameterized single-qubit rotations about the Bloch sphere axes are denoted by $R_x$, $R_y$, and $R_z$. Network outputs are collectively denoted $y$, representing the predicted output for a given input. For QNNs, $y$ is obtained from measurement expectation values, while for ANNs it is computed directly from the network forward pass.

To assess the expressive capacity of each architecture, we compute the global effective dimension for the two- and six-parameter ANNs and QNNs, with all values converging by $10^6$ data samples. The two-parameter QNN, see Fig.~\ref{fig:qnnCircuits}(a), consists of single-qubit rotations, a single CNOT entangling operation, and two variational rotation gates. The six-parameter QNN, Fig.~\ref{fig:qnnCircuits}(b), extends this ansatz by applying a sequence of three Euler-angle rotations to each qubit, interleaved around the sole CNOT gate available on Ascella. This structure mirrors universal two-qubit parameterizations commonly used in photonic variational algorithms \cite{vqePeruzzo2014}. Although the hardware restricts us to one entangling gate, even a single entangling operation should improve trainability by increasing expressivity and mitigating barren plateaus relative to separable circuits \cite{qnnAbbas2021}.

The resulting effective dimensions are summarized in Fig.~\ref{fig:qnnCircuits}(e). Higher effective dimension corresponds to greater expressive capacity and an increased ability to represent complex decision boundaries. In both parameter regimes, the QNNs achieve a higher effective dimension than their ANN counterparts, with the two-parameter QNN showing the greatest relative advantage (0.95 versus 0.68). This aligns with the observations of Abbas et al. \cite{qnnAbbas2021}, who report that appropriately chosen variational quantum circuits can outperform comparably sized classical models in expressivity, whereas poorly matched ansätze may provide little or no benefit. In our case, the six-parameter QNN exhibits only a marginal increase (of 0.01) in effective dimension over the six-parameter ANN, highlighting that deeper feature-encoding layers and longer ansätze do not always translate into greater expressivity for small-scale problem sets and models.

In all cases, quantum circuits were initialized in $\ket{0} \otimes \ket{0}$ and executed using angle-encoded inputs i.e., each normalized feature value was used as a rotation angle \cite{qnnAbbas2021}; expectation values were then estimated from repeated sampling. Effective dimension values were obtained by training each model using a cross-entropy loss function that quantifies the difference between predicted and true class labels and the Adam algorithm, a gradient-based stochastic optimizer \cite{adamKingma2014}, for efficient and stable convergence during simulation. This choice is appropriate in the noiseless (or low-noise) setting of classical simulations, where gradients can be computed accurately and economically. Later, when deploying QNNs on real photonic hardware, and for direct comparison with ANNs trained under identical optimization rules, we switch to the gradient-free Constrained Optimization BY Linear Approximation (COBYLA) algorithm \cite{cobylaPowell1994}. COBYLA is often preferable in variational quantum algorithms as it operates using local linear approximations within a constrained parameter space, avoiding explicit gradient evaluations and providing robust, efficient convergence in low-dimensional settings under shot noise and hardware imperfections \cite{qnnOptCompJarman2021}.

\begin{figure}[H]
    \centering
    \includegraphics[width=\textwidth]{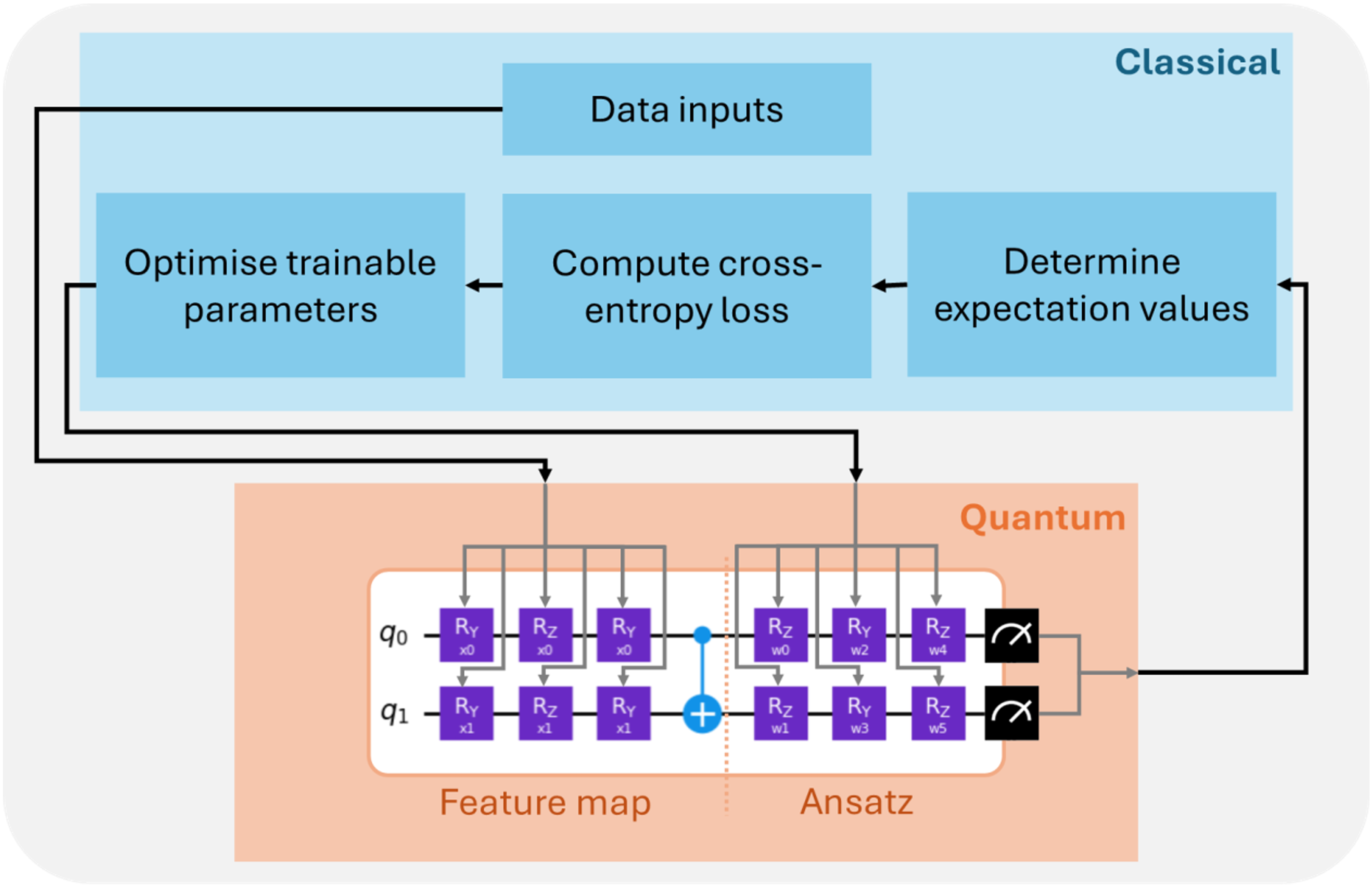}
    \caption{Structure of the hybrid quantum–classical algorithm used to train QNNs. Classical features are first normalized and encoded into the quantum circuit using angle encoding. The qubits $q_0$ and $q_1$ (initialized in the computational basis state $\ket{0}$) are then processed by the parameterized QNN (the example shown corresponds to the circuit in Fig.~\ref{fig:qnnCircuits}(b)), which is executed remotely on the quantum processor. Measurement statistics are collected to estimate output expectation values. A classical optimizer updates the circuit’s trainable parameters based on the loss function, and the procedure iterates until convergence.}
    \label{fig:hybridStructure}
\end{figure}

The broader hybrid quantum–classical training structure used for the QNNs is shown in Fig.~\ref{fig:hybridStructure}. For the classical baselines, the quantum block and expectation value computation step in Fig.~\ref{fig:hybridStructure} is replaced by the ANN architectures shown in Fig.~\ref{fig:qnnCircuits}(c) and Fig.~\ref{fig:qnnCircuits}(d), resulting in fully classical classifiers trained under the same loss and optimization scheme. This unified framework enables a direct, parameter-matched comparison of representational capacity across classical and quantum models.

\begin{figure}[H]
    \centering
    \includegraphics[width=\textwidth]{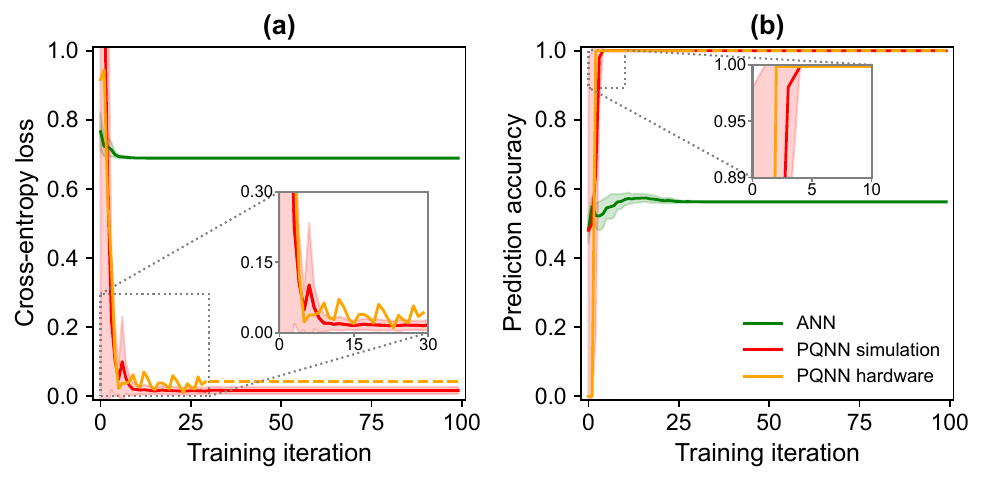}
    \caption{Online training performance on the XOR dataset for the two-parameter neural networks. (a) Cross-entropy loss and (b) prediction accuracy over training iterations. The QNN (red lines), shown in Fig.~\ref{fig:qnnCircuits}(a) to include a single entangling operation and two variational parameters, successfully learns the non-linear decision boundary, whereas the ANN (green lines) consisting of a single-layer perceptron, shown in Fig.~\ref{fig:qnnCircuits}(c), cannot. Simulated ANN and QNN curves show mean behavior with ±1 standard deviation shading. For QNNs deployed on hardware, solid orange lines show the single-run trajectory and dashed lines from iteration 31 indicate the final hardware-measured performance returned by the optimizer. The hardware run terminates earlier than the simulation due to resource limits, but convergence behavior is still observed.}
    \label{fig:xorResults}
\end{figure}

We first benchmark the two-parameter models on the XOR dataset. It is well known that a single-layer perceptron, such as the one shown in Fig.~\ref{fig:qnnCircuits}(c), cannot solve nonlinearly separable problems like XOR \cite{minXorSingh2016}. In contrast, the QNN shown in Fig.~\ref{fig:qnnCircuits}(a), despite using the same number of trainable parameters, successfully learns the correct decision boundary. This behavior is consistent with its substantially higher effective dimension listed in Fig.~\ref{fig:qnnCircuits}(e), indicating greater expressive capacity in low-parameter regimes. The minimal classical architecture capable of solving XOR requires at least one hidden layer and eight trainable parameters \cite{minXorSingh2016}, whereas our photonic QNN (PQNN) achieves perfect classification with only two. This highlights an algorithmic advantage: the quantum model requires fewer parameters (and a structurally simpler network) to represent the same function. Training performance for both models is shown in Fig.~\ref{fig:xorResults}: in all simulated curves, solid lines denote mean performance and shaded bands denote ±1 standard deviation. For hardware-executed QNNs, solid lines show the trajectory of a single training run, while dashed lines indicate the final hardware-measured performance obtained.

The QNN achieves a converged cross-entropy loss below 0.1 and 100\% prediction accuracy, whereas the single-layer ANN (perceptron) saturates near random-guessing performance with a final loss of approximately $ln(2)\approx0.69$. Although this gap demonstrates a representational advantage (consistent with the substantially higher effective dimension of the QNN), it does not constitute a runtime or computational advantage: the XOR task can be efficiently solved by a slightly larger classical model, such as a minimal multilayer perceptron with $\geq8$ trainable parameters \cite{minXorSingh2016}, and at this scale ($\lesssim100$ qubits) the quantum circuit is also classically simulable. These results therefore constitute a small-scale example of algorithmic advantage rather than full quantum advantage, as despite the QNN achieving superior performance at this fixed parameter scale, the task remains efficiently solvable by classical models of marginally higher complexity, such as multilayer perceptrons with increased parameter counts.

Nevertheless, demonstrations of algorithmic advantage are important stepping stones: as datasets become higher dimensional, classical training costs grow rapidly \cite{deepLearnNajafabadi2015}, whereas the expressive power of quantum circuits can in principle scale exponentially with circuit depth and qubit number \cite{robustQmlSpeedupLiu2021, quantumDataHuang2022}. Small-scale experiments therefore serve as critical validation points for future hardware where such scaling may become accessible.

\begin{figure}[H]
    \centering
    \includegraphics[width=\textwidth]{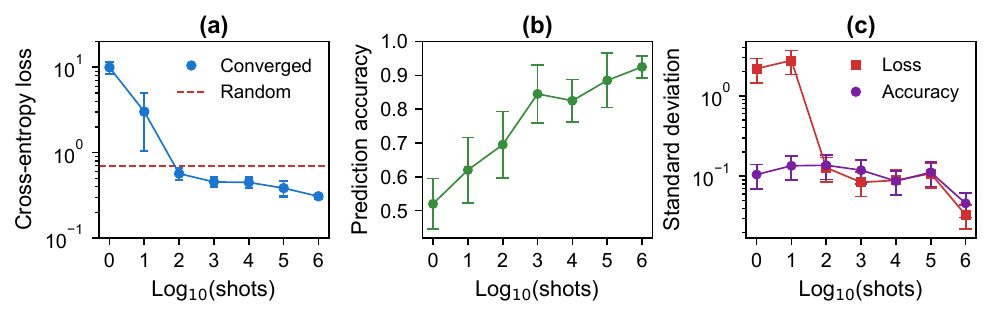}
    \caption{Simulated effect of finite shots (repeated measurements) on the performance of the six-parameter QNN trained on the Iris subset. (a) Final mean cross-entropy loss vs. shots, with the dashed horizontal line marking the random-guessing value for a balanced two-class problem, $\ln 2 \approx 0.69$. (b) Final mean accuracy vs. shots. (c) Standard deviation of converged loss and accuracy vs. shots. Error bars computed over $n=10$ independent training runs with panels (a) and (b) showing 95\% confidence intervals and panel (c) showing standard deviation across runs scaled by $\sqrt{n-1}$ to quantify variability \cite{krzywinski_error_2013}.}
    \label{fig:samplingEffects}
\end{figure}

To characterize the robustness of QNNs to finite-shot sampling, we simulate the two-parameter QNN on the Iris subset while varying the number of measurement shots per circuit evaluation. Photonic hardware parameters were chosen based on those reported for Ascella with brightness 55\%, indistinguishability 86\%, normalized second-order correlation function $g^{(2)}=0.00183$, transmittance 2.2\%, and phase shifter imprecision 1 mrad \cite{ascellaMaring2024}. To mitigate photon loss, we only count measurement shots (samples) when a photon event is detected. As shown in Fig.~\ref{fig:samplingEffects}, as expected, the final cross-entropy loss improves rapidly with increasing shot count, transitioning from random performance at tens of shots to meaningful convergence beyond roughly 100 shots. Classification accuracy shows a corresponding increase, and the variance across independent training runs decreases gradually, stabilizing at standard deviations $\lesssim0.1$ when using $\gtrsim100$ shots. For all subsequent photonic simulations and hardware runs, we use $10^5$ shots per iteration to mitigate sampling noise.

We next study the larger QNN, shown in Fig.~\ref{fig:qnnCircuits}(b),which despite having triple the trainable parameters has a lower effective dimension than the two-parameter QNN. Consistent with this, the larger QNN exhibits worse performance on the Iris subset than the more task-appropriate smaller QNN, yeilding a worse mean converged cross-entropy loss of 0.06 compared to 0.02. This reinforces the point that expressivity, not parameter count, determines trainability in variational circuits.

\begin{figure}[H]
    \centering
    \includegraphics[width=\textwidth]{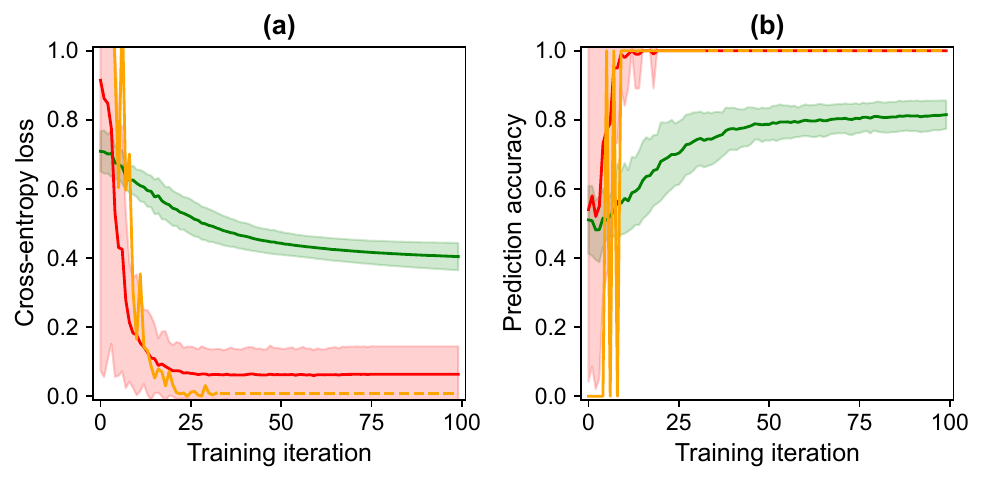}
    \includegraphics[width=\textwidth]{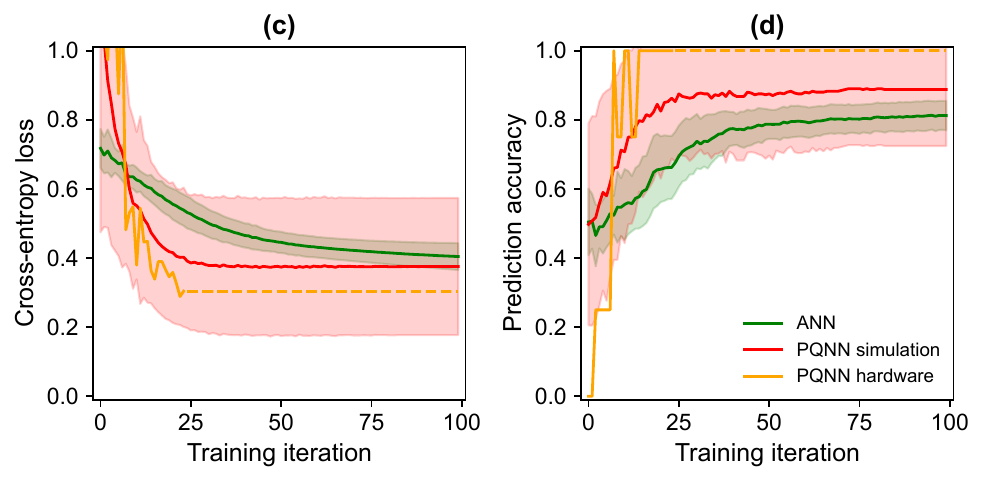}
    \caption{Online (top row) and offline (bottom row) learning of the Iris subset using six-parameter QNN and ANN. Left panels show cross-entropy loss progression; right panels show accuracy. Shaded regions denote ±1 standard deviation for simulations. Dashed lines show the final hardware-measured performance returned by the optimizer.}
    \label{fig:irisBatchComp}
\end{figure}

We also compare our default online learning (batch size = 1) to mini-batch learning (batch size = 4) in Fig.~\ref{fig:irisBatchComp}. For the larger QNN, online learning (updating the parameters after each individual training iteration) leads to substantially better convergence, reducing the converged mean loss by 0.32 and improving the converged mean accuracy by 11\%. This behavior aligns with observations in classical deep learning, where smaller batch sizes can improve convergence and generalization by providing more frequent, up-to-date updates and preserving beneficial stochasticity in the optimization process \cite{smallBatchMasters2018}. In our case, the stochastic component of single-sample updates appears to help the gradient-free COBYLA optimizer explore the loss landscape more effectively. By contrast, mini-batching averages these fluctuations, hindering convergence for these small datasets and shallow circuits ($<10$ cascaded gates) by suppressing potentially useful variability in the parameter updates.

Finally, we test the higher-ED QNN, using the same circuit deployed on two distinct quantum hardware platforms: Quandela's Ascella photonic processor and one of IBM’s Heron-R2 superconducting processors. For both platforms, results are obtained without applying error mitigation or error correction techniques. While mature mitigation tools are available for superconducting processors, analogous techniques were not as readily accessible on the photonic platform used. We therefore avoid applying mitigation selectively to one platform, in order to preserve a like-for-like comparison. In this regime, we attribute differences in training stability to primarily reflect underlying hardware noise and gate fidelity rather than the absence of platform-specific mitigation techniques.

Due to resource constraints, while the photonic runs use $10^5$ shots per iteration, the superconducting runs use $10^4$. Importantly, this difference does not qualitatively affect the comparison presented here. As shown in Fig.~\ref{fig:samplingEffects}, we operate well beyond the convergence threshold, such that increasing the number of shots further does not materially change the achieved loss or accuracy. At these shot counts, sampling noise is negligible compared to optimization variance, consistent with the trends in Fig.~\ref{fig:samplingEffects} and more generally standard shot-error scaling \cite{shotNoiseEqLiu2024}. As a result, the observed performance differences should be primarily attributed to hardware characteristics (such as gate imperfections) and their knock-on effect on optimization dynamics rather than finite-shot effects. As with the other results in this work, each hardware series plotted corresponds to a single training run, while statistical variability (indicating convergence stability) is quantified via repeated simulated runs (100 trials), shown as ±1 standard deviation shaded regions. The results are shown in Fig.~\ref{fig:sQnn}:

\begin{figure}[H]
    \centering
    \includegraphics[width=\textwidth]{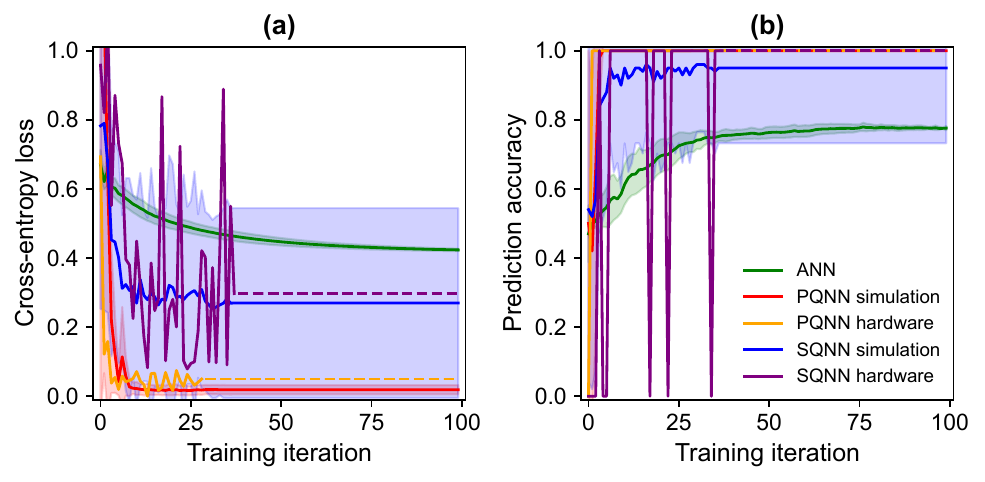}
    \caption{Online learning of the Iris subset using two-parameter ANN and QNN deployed on both photonic (PQNN) and superconducting (SQNN) hardware, respectively. (a) Cross-entropy loss and (b) accuracy vs. training iteration. Shaded regions denote ±1 standard deviation for simulations. Dashed lines show the final hardware-measured performance returned by the optimizer.}
    \label{fig:sQnn}
\end{figure}

Compared to the superconducting platform, the photonic QNN exhibits more stable convergence with lower variance across simulated runs ($<0.05$ compared to $>0.5$), consistent with Ascella’s high reported gate fidelities \cite{ascellaMaring2024}. Furthermore, its intrinsically faster gate operation further enables substantially more shots per training iteration, reducing sampling noise. While this higher shot availability can in principle further stabilize gradient estimation, its effect is secondary in the present experiments, where sampling noise is already subdominant. Regarding operation speed, assuming an optimistic 100 ns gate time (and excluding classical high-speed processing) \cite{ibmTgAmer2025}, our QNN with four gates achieves a net shot rate of only $\approx 3$ kHz on superconducting hardware, whereas on photonic hardware (80 MHz source repetition rate and assuming a conservative 2.2\% end-to-end transmittance with a CNOT success probability of 1/9 \cite{ascellaMaring2024}) a net shot rate of $\approx 196$ kHz is achieved. Thus, even in the presence of significant photon loss and a probabilistic entangling operation, the photonic implementation offers an orders-of-magnitude faster net operation time. Since the quantum circuits deployed are identical between qubit platforms, the primary advantage conferred by higher shot rates in the photonic platform here is simply improved training stability and reduced elapsed real time.

In conclusion, we present, to our knowledge, the first experimental demonstration of a gate-based variational quantum classifier using single photons, and evaluate two- and six-parameter variational circuits on two binary-classification tasks. Both photonic and superconducting QNNs outperform parameter-matched ANNs, achieving lower final cross-entropy loss ($\gtrsim 0.1$ decrease) and higher prediction accuracy ($\gtrsim 10\%$ increase), including a clear algorithmic advantage on the non-linear XOR task where the classical network fails to learn. We also explored the role of hyperparameters (batching, optimizer choice, and shot count) and quantified model expressivity via the global effective dimension. Although these models remain classically simulable, the observed performance gap suggests that expressive benefits of QNNs can already manifest at very small ($\leq 6$) parameter counts. As hardware scales in qubit number and circuit depth, this expressivity gap may translate into practical near-term quantum advantage.

Our results show that small gate-based photonic QNNs (with only 2–6 trainable parameters) can surpass classical ANNs of identical parameter count. This indicates that quantum models, realized here by photonic qubits, can access function classes that classical models of comparable size cannot. This is consistent with their larger effective dimension, as previously proposed in ref. \cite{qnnAbbas2021}, and is here demonstrated experimentally on photonic hardware for nonlinearly separable classification tasks. Whether this advantage extends to higher-dimensional inputs and deeper circuits remains to be seen once quantum hardware progresses past the capabilities of classical simulation.

Projected near-term improvements, such as a 320 MHz source repetition rate and 27\% overall transmissivity \cite{ascellaMaring2024}, would increase the net shot rate of our two-parameter PQNN to nearly 10 MHz. This intrinsic speed advantage is further amplified in practice by the larger number of shots made available on the cloud-accessible photonic hardware, enabling lower-noise gradient estimates and more stable training.

Looking ahead, photonic platforms remain strong candidates for scalable QNNs. Recent milestones, including high-fidelity two-qubit gates \cite{ascellaMaring2024} and rapid growth in dual-rail qubit capacity (doubling in under three years \cite{belenosQuandela2025}), point toward circuits of increasing size and complexity. A natural next step is to extend these demonstrations to measurement-based models, which are fully algorithmically equivalent but more hardware-scalable than gate-based photonic architectures \cite{mbqcRaussendorf2003, fbqcBartolucci2023}. Replicating our results on such cluster-state platforms would provide a strong test of whether the algorithmic benefits observed here translate to the next generation of photonic quantum processors.

\section*{Data availability}
The datasets generated and analyzed during the current study are available from the corresponding author upon reasonable request. 
The Iris dataset is publicly available from the UCI Machine Learning Repository (\url{https://doi.org/10.24432/C56C76}).

\section*{Code availability}
The code used to generate and analyze the data in this study is available from the corresponding author upon reasonable request, subject to institutional and third-party restrictions.

\section*{Acknowledgments}
The authors thank Dr Prakash Murali for useful discussions on shot-count effects and error mitigation. This work was supported in part by the Engineering and Physical Sciences Research Council (EPSRC) Centre for Doctoral Training in Photonic and Electronic Systems under Grant EP/Y034864/1 and in part by Leonardo UK Ltd.

\section*{Author contributions}
The project was conceived by S.M., who also performed the numerical simulations and experimental runs. L.S. provided theoretical input and supervision. Both authors contributed to the writing of the manuscript.

\section*{Competing interests}
S.M. was previously employed by and is currently affiliated with Leonardo UK Ltd through an industrial co-sponsored PhD; this affiliation did not influence the conduct or reporting of this work. L.S. declares no competing interests.

\bibliographystyle{naturemag} 
\bibliography{references}

\end{document}